# Towards a quantum evolutionary scheme: violating Bell's inequalities in language[1]

Diederik Aerts
Leo Apostel Centre for Interdisciplinary Studies (CLEA) and
Foundations of the Exact Sciences (FUND)
Department of Mathematics
Free University of Brussels
Krijgskundestraat 33
1160 Brussels – Belgium
diraerts@vub.ac.be

Marek Czachor
Foundations of the Exact Sciences (FUND),
Free University of Brussels, and
Faculty of Applied Physics and Mathematics,
Politechnika Gdanska
ul. Narutowicza 11/12
80-952 Gdansk – Poland
mczachor@pg.gda.pl

Bart D'Hooghe
Centre Leo Apostel for Interdisciplinary Studies (CLEA) and
Foundations of the Exact Sciences (FUND)
Department of Mathematics
Free University of Brussels
Krijgskundestraat 33
1160 Brussels – Belgium
bdhooghe@vub.ac.be

## Abstract

We show the presence of genuine quantum structures in human language. The neo-Darwinian evolutionary scheme is founded on a probability structure that satisfies the Kolmogorovian axioms, and as a consequence cannot incorporate quantum-like evolutionary change. In earlier research we revealed quantum structures in processes taking place in conceptual space. We argue that the presence of quantum structures in language and the earlier detected quantum structures in conceptual change make the neo-Darwinian evolutionary scheme strictly too limited for Evolutionary Epistemology. We sketch how we believe that evolution in a more general way should be implemented in epistemology and conceptual change, but also in biology, and how this view would lead to another relation between both biology and epistemology.

## 1. Introduction

In this article we will demonstrate the presence of quantum structures in language by proving the violation of Bell's inequalities[2]. Apart from any specific philosophical theory about language, the presence of quantum structures in language is an interesting finding, particularly to the community of scientists who study the structure of language, and also to those who investigate the working of the human mind, language being one of the mind's major products. In addition to this first-order motivation – as one could call it –, the search for quantum structures in language was equally well guided by the aim of elaborating the *global worldviews construction program*, which is the main theme of research in the Leo Apostel Centre (CLEA) (Aerts, Apostel, De Moor et al. 1995; Apostel 1995). It is through its relevance for the global worldviews construction program that the material presented in this article becomes relevant for Evolutionary Epistemology (EE), certainly in the broad interpretation that EE was given at the conference where we presented the material of this article. The Stanford Encyclopaedia of Philosophy defines EE thus:

> EE is the attempt to address questions in the theory of knowledge from an evolutionary point of view. It involves, in part, employing models and metaphors drawn from evolutionary biology in the attempt to characterise and resolve issues arising in epistemology and conceptual change. There are two interrelated but distinct programs which go by the name EE. One focuses on the development of cognitive mechanisms in animals and humans. This involves as straightforward extension of the biological theory of evolution to those aspects or traits of animals which are biological substrates of cognitive activity, *e.g.* their brains, sensory systems, motor systems, etc. This program has been labelled EEM. The other program attempts to account for the evolution of ideas, scientific theories, epistemic norms and culture in general by using models and metaphors drawn from evolutionary biology, and it has been labelled EET.

EE is sometimes given a narrow meaning, according to which it is said to attempt to model epistemology not merely in terms of biological evolution but rather in terms of the neo-Darwinian version of biological evolution, basing itself on the assumption that it is this neo-Darwinian version that can be transferred to epistemology. The broad meaning of EE is that the inspiration for modelling epistemology remains biological evolution, but the neo-Darwinian form of this biological evolution is considered to be just one of the possibilities.

Within the framework of our *global worldviews construction program* at CLEA we have introduced the model of *the layered structure of reality*. This model assumes that the existence of separate bodies of knowledge, *i.e.* scientific disciplines, concerning parts or fragments of reality has its reasons. The disciplines in the theoretical sciences have not been chosen in an arbitrary way but correspond to a layered organisation of reality itself. Reality indeed shows itself to us in the form of layers. The first is a pre-material layer of elementary particles and waves that are the subject matter of quantum physics. This is followed by a material layer, studied by physics and chemistry. The next layers are those of life forms that make up the field of study of biology, and of interacting life forms studied by sociology. Finally, we have the psycho-cognitive layer, studied by psychology and cognitive science. This layered structure is considered effective and real and not merely a suitable classification. On the other hand, it is clear that the different layers are not separated, but in constant interaction and connected in all kinds of ways, *i.e.* through contextual, emergent, and downward causation influences. They are *forms of condensation* in reality as an undivided totality. The study of this multi-layered structure of reality, of how the different layers are interconnected, how they emerge one from another, et cetera, is one of the encompassing research themes in CLEA.

---
[2] Bell-inequalities are a set of mathematical inequalities that have been derived with the aim of testing experimentally the presence of quantum structures within the data structure of an arbitrary source (see section 2.4).

The way in which the *psycho-cognitive layer* of reality is considered a part of this layered structure shows that we advocate a naturalised epistemology within this approach. Epistemology is considered to be one of the processes that take place within the psycho-cognitive layer of reality. As such, this layered structure is considered to have grown under the influence of evolution, so that the approach followed at CLEA is compatible with the broad meaning of EE. As we will see in the following, the detection of quantum structures in the psycho-cognitive layer, more specifically in language, makes it possible for us to formulate a criticism of the narrow meaning of EE. The neo-Darwinian view on evolution is in fact a classical view, also if we use the word classical as opposed to quantum, *i.e.* in the way it is used by physicists. The reason is that the two basic mechanisms of neo-Darwinian evolution, namely *variation* and *selection*, on which the whole view is founded, turn the resulting evolution mechanism into a classical mechanics mechanism. The identification of quantum structures in what we have called the psycho-cognitive layer proves the neo-Darwinian scheme to be too limited for modelling evolution in this layer. To explain in a more concrete way why and how the neo-Darwinian scheme is too limited, we need to introduce new concepts and explain results of foregoing research. We will do this along with an exposition of different but compatible findings of quantum structures in the psycho-cognitive layer in the sections to come.

## 2. Quantum vs classical probability

To clarify the nature of the shortcoming of the neo-Darwinian scheme for evolution that we intend to reveal, we need to explain the difference between classical probability and quantum probability.

### *2.1. The nature of classical probability*

Microscopic physical particles are described by quantum mechanics, which is an indeterministic theory. This means that for a quantum experiment the quantum entity can be prepared in a state such that the outcome of the measurement is not predictable with certainty. Of course, indeterminism of a specific set of experimental setups is not strictly reserved to the micro-world, since indeterminism as a phenomenon related to a specific experiment or set of experiments occurs regularly in the macro-world as well, *e.g.* the problem of forecasting the weather, which is studied in chaos and complexity theories. Another example is how in the field of population dynamics (Hoppenstedt 1982) change is described by Markovian chain processes (Markov 1906; Doob 1953) using discrete time such that at each moment the state of the system is completely defined by the previous state. Hence classical indeterminism emerges in situations in which *the observer lacks knowledge about the specific state of an underlying deterministic world*. In a classical world, no indeterminism is involved on this deepest level of reality.

### *2.2. Hidden variable theories for quantum mechanics*

The structure of indeterminism encountered in quantum mechanics is of a completely different and new nature which has, at first sight at least, no counterpart in the macro-world. John von Neumann (von Neumann 1932) proved that a hypothetical classical underlying theory yielding the probabilities of quantum mechanics as due to a lack of knowledge of the state would never be able to produce the correct numerical probabilities as the ones encountered in quantum mechanics. The hypothetical classical theories considered by von Neumann have been called *hidden variable theories*, because they introduce extra variables such that when all these variables are known every outcome of every experiment can be predicted, and hence the theory is classical deterministic. However, these extra variables are *hidden*, which means that we do lack knowledge on their values, and it is this lack of knowledge that gives rise to probabilities, and hence a classical type of probability.

John Bell remarked that one of the assumptions[3] made by von Neumann does not need to be satisfied for any hidden variable theory, and Bell built on the spot a hidden variable model for a specific quantum system (Bell 1966). However, Bell's hidden variable model was a very theoretical model, making it impossible to establish the physical meaning of the hidden variable and hence to determine the *wrong* assumption in von Neumann's no-go theorem.

*2.3. Classical versus quantum probabilities: the structural difference*

The different nature of quantum and classical indeterminism is expressed by a deep structural difference of a mathematical nature between the probability model of classical mechanics and that of quantum mechanics. Classical probability theory was axiomatised and elaborated into a formal mathematical theory by Kolmogorov in 1933 (Kolmogorov 1950). The probability model that appears in situations described by classical mechanics – in which the probability is due to a lack of knowledge about the state of the physical entity under consideration – satisfies these axioms of Kolmogorov, and it is called a Kolmogorovian probability model. The probability model that appears in quantum mechanics does not satisfy the axioms of Kolmogorov (Foulis and Randall 1972; Accardi 1984; Gudder 1988; Pitowsky 1989), such that quantum probability cannot be explained as being due to a lack of knowledge about the state of the system. It is this non-Kolmogorovian nature of the quantum probability model that was already at the core of the proof of John von Neumann's no-go theorem. This was to appear with greater emphasis yet in later corroborations of von Neumann's no-go theorem (Jauch and Piron 1963; Kochen and Specker 1967; Gudder 1968).

*2.4. Bell's inequalities and Kolmogorovian probability models*

In 1964, Bell analyzed the physical situation presented in the Einstein Podolsky Rosen paper (Einstein, Podolsky and Rosen 1935) and derived the first formulation of what we now know as *Bell's inequalities* (Bell 1964). Bell's inequalities are a set of mathematical inequalities formulated by means of expectation values of outcomes of experiments. Bell proves that when there is a *local* hidden variable model for the considered quantum entity, the expectation values predicted for the considered experiments and appearing in Bell's inequalities will be such that the inequalities are *not* violated. By contrast, quantum mechanics predicts that the inequalities *will* be violated. All of the many experiments that were executed yielded highly convincing evidence of Bell's inequalities being violated in the very way predicted by quantum mechanics (Clauser and Shimony 1978; Aspect 1999). By introducing the violation versus non-violation of Bell's inequalities principle, Bell offered an operational manner of establishing whether the probability model of a system could be explained from a lack of knowledge about an underlying deterministic world. Pitowsky proved that a probability model is Kolmogorovian if and only if none of the Bell-type inequalities that can be defined by means of expectation values of correlations between different joint measurements of the relevant set of measurements are violated. This means that the detection of a single Bell's inequality being violated suffices to render the collection of probabilities incompatible with a Kolmogorovian structure (Pitowsky 1989). Before that, Luigi Accardi (Accardi 1984) had already proved a similar result in a different setting, which was, however, less easy to generalise.

---

[3] The technical assumption that the expectation value of a linear combination of two observable quantities equals the linear combination of the expectation values.

*2.5. The violation of Bell's inequalities in different layers of reality*

A situation giving rise to the violation of Bell's inequalities with measurements on a macroscopic physical entity was considered in Aerts (1982). This model was proven to entail a non-Kolmogorovian probability model, albeit that its non-Kolmogorovian aspect derived from the introduction of a *classical lack of knowledge probability* into the measuring apparatuses used in the example. The contextually driven mechanism identified as causing the probability model to acquire a non-Kolmogorovian structure shows two necessary effects: (1) the interaction with the measurement context changes the state of the system; and (2) there is a lack of knowledge on the way these measurements influence the state of the mechanical system, and hence on the interaction between the measurement context and the system. The existence of a *lack of knowledge on measurements that change the state of the entity under study* is abundantly present in many macroscopic situations (Aerts 1982, 1986; Czachor 1992). Much more so than in cases of macroscopic mechanical systems involving measurements that show these two effects, this type of situation will occur naturally with entities that are the common subjects of scientific disciplines in what we know as the human sciences, such as psychology and sociology, since both these effects are typical of the majority of measurements carried out in these fields. This is why it is plausible that non-Kolmogorovian probability structures will be encountered in those layers of reality where the two above effects are present, more specifically in the psycho-cognitive layer of reality.

**3. Quantum structures and the human mind**

In this section we will discuss the different ways in which we have identified quantum-like structures in the psycho-cognitive layer, where the (measurement) context is crucial.

*3.1. Modelling decision processes*

The first situation that we analysed in this way is that of psychological decision processes, where subjects are influenced by, and form part of their opinions during the testing process (Aerts and Aerts 1994, 1997; Aerts 1995). We have set out to prove that the probability model encountered in these situations is non-Kolmogorovian. Let us consider the simple situation of an opinion poll, because here the cause of the non-Kolmogorovian structure can be seen intuitively. Let us suppose the survey contains the following question: "Are you in favour or against the legalisation of soft drugs?" The respondents will comprise (1) those that were already in favour before being asked the question; (2) those that were already against before they are asked; and (3) those that were neither in favour nor against before being asked, and make up their opinion in the course of the survey. It is the state of mind of the respondents in the latter group that will be affected by the measurement context (the manner in which the survey is conducted, but also all the details about the environmental situation during the survey), while moreover we lack knowledge on the exact nature of this change of state. These are the two effects that turn the probability model into a non-Kolmogorovian model (Aerts and Aerts 1994, 1997; Aerts 1995).

We should point out the following here. Social scientists studying situations that involve opinion polls are of course very much aware of the existence of a subgroup of respondents that have *no opinion* before they are asked to answer the questions. This is usually resolved by including the answer *no opinion* or *don't know*, next to *in favour* and *against*. A quantum model behaves differently. In a quantum model, the introduction of such a third outcome – comparable to the outcome *no opinion* – does not offer an adequate solution. There will still be a group that does not fit in with any of the three alternatives offered, because they do not allow classification along these lines prior to the survey. These respondents did not decide to vote for the third *no opinion* outcome (which would make them have an opinion after all). Their mind is literally *made up* during the survey itself, and hence classifying them as having

*no opinion* prior to the survey would be equally erroneous. There is another aspect we wish to point out. Although ad hoc models are made in the field of social sciences for these situations, all these models extract the statistical analysis that they use from *classical statistics*. Classical statistics is a mathematical theory that is built on the mathematical foundation of a Kolmogorovian probability model. Stating that the probability model involved is non-Kolmogorovian is equal to stating that *classical statistics does not apply* in this situation.

### 3.2. Modelling concepts

If decision processes entail a quantum-like structure it is plausible that other structures directly connected to the functioning of the human mind equally contain quantum structure. In this respect we were triggered to investigate one of the old and unsolved problems in *concept research*, called the *pet-fish problem*. The pet-fish problem is encountered in concept research when mathematical models for single concepts (pet and fish in this case) are tempted to be combined mathematically to give rise to a model for the combined concept (pet-fish in this case). How to do this is a long standing open problem in concept research. We were able to solve the pet-fish problem by explicitly using a quantum mechanical mathematical representation for concepts (Aerts and Gabora in press/ab). Let us sketch in more detail the nature of this solution.

Following the *classical theory* of semantic concepts, there is for each concept a set of defining properties that are necessary and sufficient for category membership. Essential shortcomings of this theory have been identified in many ways (Smith and Medin 1981; Komatsu 1992). A fundamental step forward was taken in the seventies, under the influence of the work of Eleanor Rosch and her collaborators at Berkeley. They showed that the *typicality* of exemplars of a concept and the *application values* of properties vary. For example, an *apple*, which is an exemplar of the concept *fruit* is a more typical exemplar than a *pineapple*, and *juicy* has a higher application value as a property of the concept *fruit* than the property *expensive* has for the concept *fruit*. Rosch formulated the prototype theory: each concept has a prototype and the typicality of an exemplar depends on how similar it is to the prototype (Rosch 1975, 1978, 1983; Rosch and Mervis 1975). The breakthrough was that a concept appears as a fuzzy structure, where similarity becomes the measure. A variety of models were proposed, and next to the ones inspired by the prototype idea, the exemplar theories became a second important class. In this category it is not the prototype but salient exemplars that serve as reference; typicality and application value are measured as the distance to these exemplars (Nosofsky 1988, 1992; Heit and Barsalou 1996). These theories yield satisfactory predictions in the case of experiments with single concepts for many dependent variables, including typicality ratings[4] and exemplar generation frequencies [5]. However, problems arise when it comes to *combinations of concepts*, for these theories do not allow to account for phenomena such as the so-called *guppy effect*, where guppy is not rated as a good example of *pet* nor of *fish*, but it is rated as a good example of *pet-fish* (Osherson and Smith 1981). General fuzzy set theory has been tried in vain to deliver a description of the *guppy effect* (Osherson and Smith 1982; Zadeh 1982), and this peculiarity can also be understood intuitively: if (1) activation of *pet* causes a small activation of *guppy*, and (2) activation of *fish* causes a small activation of *guppy*, how is it that (3) activation of *pet-fish* causes a large activation of *guppy*?

---

[4] Typicality ratings are ratings made by the subject to whom the experimenter proposes a collection of different exemplars of the same concept, and a collection of different application values of properties of that concept. The ratings can be made in different ways; often a scale from 1 to 7 is used.

[5] Exemplar generation frequencies correspond to tests where subjects are asked to generate exemplars of a specific concept. The frequency of a specific exemplar is considered to be correlated with the typicality of that exemplar.

The combination problem is considered so serious that it has been said that not much progress will be possible in the field as long as no light is shed on this problem (Fodor 1994; Kamp and Partee 1995; Rips 1995; Hampton 1997). As a consequence, existing theories concentrate on attempts to model the combination of at most two concepts (*red apple*, *fake diamond*, *car-seat*, *brain-storm*, *pet-fish*, *etc.*), while the real challenge consists in modelling a sentence, or a set of sentences. Our *quantum mechanical* theory for concepts models an arbitrary combination of concepts, including combinations consisting of more than two concepts, and it also describes the *guppy effect* by making use of the standard quantum mechanical procedure to describe the combinations of quantum entities. We show the quantum effect called entanglement to be at the origin of the guppy effect (Aerts and Gabora in press/ab; Gabora and Aerts 2002a,b). Entanglement is one of the characteristic properties of quantum entities. If two quantum entities are entangled it means that a change of state of one of the quantum entities provokes a corresponding change of state of the other quantum entity. It can be proven that if there would be a physical process carrying this *change of state correlation*, then this physical process is neither a causal nor a local process. In the quantum model for concepts that we worked out, different concepts in a combination (*e.g.* a sentence) of these concepts are entangled in the sense that if one of the concepts collapses to one of its exemplars (changes its state), also the other concept undergoes a change of state. We refer to section 4 where we consider the sentence *The pet eats the food*, and where in the process of violating Bell's inequalities we consider different situations where it can be seen how a change of state of *pet* induces a change of state of *food* and vice versa, *pet* and *food* being two entangled concepts within the sentence *The pet eats the food*. The guppy effect is due to the fact that, as a consequence of entanglement, the contextual influence on one of the concepts of a combination of concepts also influences the other concepts in this combination. Literally, if in the combination *pet-fish*, the pet becomes a guppy, then also the fish becomes a guppy, because pet and fish are entangled in the combination *pet-fish*.

*3.3. Concepts and sentences versus sources and data*

The main impact of Bell's work was that it shifted the analysis of hidden variables from philosophical speculations to experiment. Bell's inequalities deliver a criterion that allows investigating the probabilistic and *logical* structure of a *source* on the basis of the data it produces. The data are represented by sequences of symbols and are as *ordinary* or *classical* as the characters used to write this article. Bell's inequalities constitute a statistical test investigating correlations between different groups of symbols, and the probability model employed in the analysis is the usual model based on the frequencies of occurrence of certain results. However, and this is one of the ingenious elements of the Bell analysis, certain frequencies are impossible if the source is characterised by a Kolmogorovian model of probability, or – which is roughly equivalent – by a Boolean logic[6]. Again, it should be borne in mind that the goal of the Bell statistical test is to reveal a non-Kolmogorovian *hidden* structure of a source.

The above situation is strikingly similar to that encountered in the investigation of logical structures behind language. A speaker or author is a source, the structures that are the subject of investigation refer to the conceptual level and are as hidden as von Neumann's hidden variables, but the data produced by the source are collections of symbols contained in a computer memory or on a piece of paper. Quantitative linguistics is the field to look for theoretical tools.

---

[6] Boolean logic is a synonym for classical logic. It is a kind of logic where the propositions form the mathematical structure of a Boolean algebra, carrying the traditional logical operations of conjunction, disjunction, complement and their combinations and formulas.

When working on the modelling of concepts, and with our approach having allowed us to model an entire sentence as a combination of the concepts contained in this sentence (Aerts and Gabora in press/ab), we stumbled rather by accident on articles on the World Wide Web proposing mathematical models for text fragments, more specifically, articles on *Latent Semantic Analysis* (LSA)[7], many of which have been made available through the web by Thomas Landauer's team at the university of Boulder (http://lsa.colorado.edu/). We saw that the mathematical models used were vector space models, which was quite a surprise, since our quantum model is also a vector space model (a Hilbert space, the principal mathematical structure used in quantum mechanics, is a vector space).

Meanwhile, we know that there are different approaches to modelling language on mathematical models, and that most of them use vector spaces as principal mathematical entities. The prominent examples of such approaches are *Latent Semantic Analysis* (LSA) (Deerwester, Dumais, Furnas, et al. 1990; Landauer, Foltz and Laham 1998), *Hyperspace Analogue to Language* (HAL) (Lund and Burgess 1996), *Probabilistic Latent Semantic Analysis* (pLSA) (Hofmann 1999), *Latent Dirichlet Allocation* (Blei, Ng and Jordan Michael 2003) or *Topic Model* (Griffiths and Steyvers 2004). These methods are based on text co-occurrence matrices and data-analysis techniques employing singular value decomposition[8]. For example, LSA provides a powerful method for determining similarity of meaning of words and passages by analysis of large text corpora. Quite impressive results have been obtained from experiments simulating human performance. For example, LSA-programmed machines proved capable of passing multiple-choice exams such as a Test of English as a Foreign Language (TOEFL) – (after receiving training in general English) (Landauer and Dumais 1997) and, after learning from an introductory psychology textbook, a final exam for psychology students (Landauer 2002).

In (Aerts and Czachor 2004) we explored the structural similarities between these vector space models for language and the Hilbert space model of quantum mechanics. Similarities of mathematical structures between quantum mechanics, and quantum logic and quantum information theory in particular, on the one hand, and those employed in LSA on the other, opened new perspectives for both LSA and the traditional quantum fields of interest. Certain structures quite natural in the quantum domains might help to clarify the difficulties of LSA. One of the first links we noticed was the *bag-of-words problem* of LSA. In LSA a text passage is treated as a *bag of words*, a set where order is irrelevant (Landauer, Laham and Foltz 1998). This is considered to be a serious problem because even on the intuitive level it will be clear that syntax is important to evaluating the meaning of text. The sentences *Mary hits John* and *John hits Mary* cannot be distinguished by LSA. The problem is basically how to *order* words in using vector models. In quantum theories, the problem was solved a long time ago in terms of tensor structures[9]. Quantum experiments confirming the violation of Bell's inequalities in fact aimed to test the presence of tensor structures in the non-Kolmogorovian model of quantum mechanical probability. In principle, analogous analyses performed on text corpora might reveal non-classical structures in their sources, *i.e.* the minds of their authors. The idea may seem exotic, but it is not that far from other mathematical techniques employed in quantitative linguistics, and whose roots are in physics. One can mention here the Zipf-Mandelbrot criterion[10] for natural languages and its generalisations resulting from non-extensive thermodynamics[11] and applied to Shakespeare's writings in (Montemurro 2001) and (Czachor and Naudts 2002).

---

[7] In LSA, texts are mapped into co-occurrence matrices. Columns and rows of the matrices represent sentences and words, respectively. Scalar products between word-vectors *(i.e.* rows of text matrices) measure similarity of meaning. LSA recognises that two different words have similar meanings. The procedure is automatic. Computers can *learn* by themselves and pass multiple-choice exams without any real understanding of the test material. Search engines based on LSA can find texts related to a given term even if this term does not occur in the text.

[8] Singular value decomposition is a representation of a matrix in the form of a product of three matrices. The middle matrix is diagonal and the elements on the diagonal are termed the singular values. For example, in applications to picture analysis, one cannot practically distinguish between two pictures whose singular values differ by small numbers. Therefore, one of the ways of compressing pictures is by setting small diagonal values to zero and keeping only the values that are sufficiently large. *Sufficient* here is determined by the intended *sharpness* of the picture. The same idea can be applied to any data that is stored as matrices – texts, for example.

Still, the possible influences go in both directions. The tools and experiences developed within LSA might open new perspectives on semantics of quantum programming languages, the languages necessary for programming quantum computers. Moreover, once we realised that tensor structures might form a common basis for quantum information and logic on the one hand, and semantic analysis on the other, we found links to still another field that had been developing independently during the past 15 years or so: distributed representations of cognitive structures in symbolic AI and neural networks[12]. To our surprise, we realised that tensor products were already employed in certain AI investigations (Smolensky 1990), and that certain alternatives to tensors had been extensively investigated in this context (Plate 2003). These structures seem unknown to the quantum information community and might prove to be of crucial importance in quantum memory models. We also mention that Widdows and Peters (2003) have investigated connections between LSA and quantum logical structures.

At the moment we are at a crossroads of at least three highly and independently developed fields of knowledge: Quantum information, semantic analysis, and symbolic AI. We are convinced that quantum probabilistic and logical structures will prove essential in the latter two domains, and it is even difficult to imagine in what direction further development will continue.

## 4. The violation of Bell's inequalities in language and its implications for EE

In this section we will present an example of how Bell's inequalities are violated if we apply the model developed for the representation of concepts to language (Aerts and Gabora, in press/ab). From what we explained in chapter 2, this proves that language contains genuine quantum structure.

### *4.1. Concepts as entities in different states under different contexts*

According to Rosch, the typicality of different exemplars of one and the same concept varies (Rosch 1975, 1978, 1983; Rosch and Mervis 1975). Subjects rate different typicalities for exemplars of the concept *fruit*, for example, resulting in a classification with decreasing typicality, such as *apple, strawberry, plum, pineapple, fig, olive*. According to our theory of concepts, *for each exemplar alone* the typicality of varies with the context that influences it (Aerts and Gabora, in press/ab). By analogy, *for each property alone*, the application value varies with the context. We performed an experiment to point out and measure our typicality and application value effect. Participants in the experiment classified exemplars of the concept *pet* under different contexts resulting in typicality ratings given in Table 1.

---

[9] Tensor structures are a generalisation of ordinary multiplication that can be applied not only to numbers but also to vectors, matrices, etc. In vector models of probability and logic, the tensor product represents logical conjunction (AND). Tensor structures are all the structures that can be obtained by tensor products of more primitive objects.

[10] In a text we can count the number of times a given word occurs, and repeat this procedure for all the words in the text. We can then produce a decreasing curve representing the frequency of occurrence of all the words, starting with the most frequent one. The Zipf-Mandelbrot criterion is a condition on the shape of this curve that allows identifying a text written in a natural language.

[11] Non-extensive thermodynamics is a statistical method of finding probability distributions that are optimal in some sense and characteristic of finite sets of data with long-range correlations. Texts written in natural languages possess such correlations due to syntax, grammar and other linguistic features.

[12] Distributed representations are models that take into account the need for information to be represented in a pattern of activation over a set of neurons, in which each concept is represented by activation over multiple neurons, and each neuron participates in the representation of multiple concepts.

| Exemplar | *The pet is chewing a bone* | *The pet is being taught* | *Look what a pet he has, I knew he was a weird person* |
|---|---|---|---|
| rabbit | 0.07 | 2.52 | 1.77 |
| cat | 3.96 | 4.80 | 0.94 |
| mouse | 0.74 | 2.27 | 3.31 |
| bird | 0.42 | 3.06 | 1.41 |
| parrot | 0.53 | 5.80 | 1.57 |
| goldfish | 0.12 | 0.69 | 0.83 |
| hamster | 0.85 | 2.72 | 1.25 |
| canary | 0.26 | 2.73 | 0.86 |
| guppy | 0.14 | 0.68 | 0.83 |
| snake | 0.57 | 0.98 | 5.64 |
| spider | 0.26 | 0.40 | 5.96 |
| dog | 6.81 | 6.78 | 0.91 |
| hedgehog | 0.53 | 0.85 | 3.48 |
| guinea pig | 0.58 | 2.63 | 1.31 |

**Table 1 Typicality ratings of different exemplars for different contexts.**

The context *the pet is chewing a bone* results in a classification with certain typicality ratings which changes when another context is applied, *e.g.* the context *the pet is being taught*. It changes again for the context *look what a pet he has, I knew he was a weird person*. The effect was also measured for the application value of a property, as can be seen in table 4 of Aerts and Gabora (in press/ab). The main idea of our concepts theory is to describe this typicality and application value effect by introducing the notion of *state of a concept*, and hence to consider a concept as an entity that can be in different states, such that a context will provoke a change in the state of the concept (Aerts and Gabora, in press/ab). Concretely, the state of the concept *pet* in the context *the pet is chewing a bone* is different from the state of *pet* in the context *look what kind of pet he has, I knew he was a weird person*. It is this *being in different states* that gives rise to the differences in values for the typicalities of different exemplars and applications of different properties. It is the set of these states and the dynamics of change of state under the influence of context corresponding to experimental data that is modelled by our quantum mechanical formalism in Hilbert space. The problem of the *combination of concepts* is resolved in our theory because in *combination*, the concepts are in different states; for example, in the combination *pet-fish*, the concept *pet* is in a state under the context *the pet is a fish*, while the concept *fish* is in a state under the context *the fish is a pet*. The states of *pet* and *fish* under these contexts have different typicalities, which explains the guppy effect (Aerts and Gabora, in press/ab).

*4.2. Two pets that eat their favourite food and violate Bell's inequalities*

Rather than begin by presenting a theoretical exposition of Bell's inequalities, we will introduce the inequalities along with the example discussed, and point out where they are violated. Again, our aim is to prove the presence of quantum structure in language.

Let us assume the following overall context. Amy and Carol, two sisters, both have a pet. Carol has a cat called Felix, and Amy has a dog with the name Roller. The cat and the dog live together with the two sisters, and get along well. They even do not mind to eat together in the same room. But of course, they eat different food, and they both are somewhat special, in that they both have one unique food brand that they prefer above all the rest. For Felix this is *Eukanuba*, a well-known brand of cat food, while for Roller this is *Royal Canin*, a famous brand of dog food. This means that in practice whenever Felix eats, she eats Eukanuba food, and whenever Roller eats, he eats Royal Canin. For a reason that is not completely clear to Amy and Carol they never touch each others' food. One more aspect, Amy and Carol can distinguish very well the food that is served in the room where the eating happens, because Eukanuba and Royal Canin have completely different smells.

Amy and Carol are both playing outside in the garden. The feeding room for the cat and dog is a room that opens onto the garden, but they are playing in a part of the garden where they cannot really see what is happening inside the room. They are however aware that one of the pets is being fed by their mom.

We will now take the sentence *The pet eats the food* as our conceptual entity, and denote it by *p*. The sentence is a *combination of concepts*, i.e. the three concepts *pet*, *eat* and *food*. The contexts considered are the following:

*e*: Hey, I think it is Roller who is eating, because I saw him just get in. (pronounced by Amy)

*f*: I believe that the food that mom served is Eukanuba, because I think I smell it. (pronounced by Carol)

*g*: One of our pets is eating one of the foods. (thought by both)

The contexts *e* and *f* are genuine contexts, *i.e.* they affect the state of the conceptual entity *The pet eats the food* if applied to it. More specifically, context *e* affects the concept *pet*, for if we write (*e*, *p*) or, in words: *Hey, I think it is Roller who is eating, because I saw him just get in. The pet eats the food*, the concept *pet* of the sentence *The pet eats the food*, is changed into *Roller*, and the sentence becomes *Roller eats the food*. Because of the overall contexts, the concept *food* in the sentence *The pet eats the food* will be affected too. Indeed, because Roller only eats *Royal Canin*, the concept *food* changes to *Royal Canin*, and the sentence *The pet eats the food* changes to *Roller eats Royal Canin*.

In a similar way, we can consider (*p*, *f*) or, in words: *The pet eats the food. I believe that the food that mom served is Eukanuba, because I think I smell it*. The context *f* affects the concept *food* of the sentence *The pet eats the food*, in the sense that *food* changes to *Eukanuba*. Hence the sentence *The pet eats the food* changes to *The pet eats Eukanuba*. Because of the overall context, the concept *pet* is also affected by the context *f*. The concept *pet* changes to *Felix*, because it is only Felix who eats Eukanuba. Hence the sentence *The pet eats the food* changes to *Felix eats Eukanuba*, under the influence of context *f*.

Context *g* is rather a trivial context. Both girls know that *one of the pets is being fed with one of the foods*. This means that it affects neither the concept *pet* nor the concept *food* nor the sentence *the pet eats the food*.

To introduce Bell's inequalities into our discussion, we will have to associate numbers with effects of different contexts. For this purpose, we will consider the effects of the contexts *e*, *f* and *g* on the sentence *p*, and define E(*e*, *p*) = +1, if it is Roller who is eating, while E(*e*, *p*) = -1, if it is Felix who is eating. Furthermore, we define E(*p*, *f*) = +1, if the food eaten is Eukanuba, and E(*p*, *f*) = -1, if the food eaten is Royal Canin. Similarly, E(*g*, *p*) = +1, if one of the foods is eaten by one of the pets, and E(*g, p*) = -1, if it is not so that one of the foods is eaten by one of the pets. Lastly, E(*p, g*) = +1, if one of the foods is eaten by one of the pets, and E(*p, g*) = -1, if it is not so that one of the foods is eaten by one of the pets.

Bell's inequalities come into play when we consider situations that vary with changing pairs of contexts for the sentence *The pet eats the food*, with one of the pairs of contexts affecting the concept *pet* in sentence *p*, and the other affecting the concept *food* in sentence *p*. More specifically, it is the four following combinations of pairs of contexts together with sentence *p* that is considered in Bell's inequalities.

(1) The pair of contexts *e* and *f* with *p*, such that *e* affects *pet* and *f* affects *food*. This is represented in symbols as (*e, p, f*) and in words as *Hey, I think it is Roller who is eating, because I saw him just get in. The pet eats the food. I believe that the food that mom served is Eukanuba, because I think I smell it.*

(2) The pair of contexts *e* and *g* with *p*, such that *e* affects *pet* and *g* affects *food*. This is represented in symbols as (*e, p, g*) and in words as *Hey, I think it is Roller who is eating, because I saw him just get in. The pet eats the food. One of our pets is eating one of the foods.*

(3) The pair of context *g* and *f* with *p*, such that *g* affects *pet* and *f* affects *food*. This is represented in symbols as (*g, p, f*) and in words as *One of our pets is eating one of the foods. The pet eats the food. I believe that the food that mom served is Eukanuba, because I think I smell it.*

(4) The pair of contexts *g* and *g* on *p*, such that *g* affects *pet* and *g* affects *food*. This is represented in symbols as (*g, p, g*) and in words as *One of our pets is eating one of the foods. The pet eats the food. One of our pets is eating one of the foods.*

We will now have to associate numbers with the effects of the contexts on these four combinations. Thus, we define E(*e, p, f*) = +1 if it is Roller who eats Eukanuba, or if it is Felix who eats Royal Canin. Similarly, E(*e, p, f*) = -1 if it is Roller who eats Royal Canin or if it is Felix who eats Eukanuba. We define E(*e, p, g*) = +1 if it is Roller who eats one of the foods, and E(*e, p, g*) = -1 if it is Felix who eats one of the foods. E(*g, p, f*) = +1 if it is one of the pets who eats Eukanuba, and E(*g, p, f*) = -1 if it is one of the pets who eats Royal Canin. Lastly, we have E(*g, p, g*) = +1 if one of the pets eats one of the foods, and E(*g, p, g*) = -1 if it is not so that one of the pets eats one of the foods.

Bell's inequalities are the following:

|E(*e, p, f*) - E(*e, p, g*)| + | E(*g, p, f*) + E(*g, p, g*)| ≤ +2

Bell's inequalities are violated whenever we have:

|E(*e, p, f*) - E(*e, p, g*)| + | E(*g, p, f*) + E(*g, p, g*)| > +2

Let us see what this gives in our situation.
(1) Case (*e, p, f*)
Amy seems to believe that it is Roller who is eating, but Carol believes that the food is Eukanuba. This means that there are three possibilities. (A) Roller is eating Royal Canin and hence Carol is mistaken about the food. (B) Felix is eating Eukanuba, and hence Amy is mistaken about which pet is eating. (C) Perhaps a new uncommon event occurred, and Roller is eating the cat food. In case A and case B, we have E(*e, p, f*) = -1. In case C, we have E(*e, p, f*) = +1.

(2) Case (*e, p, g*)
Amy believes that Roller is eating, and since the context is *g*, there is no reason to doubt her. Certainly one of the pets is eating one of the foods. Hence E(*e, p, g*) = (+1)

(3) Case (*g*, *p*, *f*)

Carol believes that the food is Eukanuba. In this case too there is no reason to doubt her. Certainly one of the pets is eating one of the foods. Hence E(*g*, *p*, *f*) = +1.

(4) Case (*g*, *p*, *g*)

One of the pets is eating one of the foods, hence E(*g*, *p*, *g*) = +1.

Let us suppose that it is certain that we are dealing with either case A or case B (thus ruling out the option of Roller's having made a move to eat cat food). In this case we have:

|E(*e*, *p*, *f*) - E(*e*, *p*, *g*)| + | E(*g*, *p*, *f*) + E(*g*, *p*, *g*)| = |-1 -1| + |+1 +1| = +4

*i.e.* an extreme violation of Bell's inequalities (indeed, it is the maximum violation).

The more we allow the possibility of case C happening, the less will be the extent of the violation of Bell's inequalities. Yet a real experiment in which the participants are presented with a situation in which the context is explained to them and they are asked to rate their assessments of what is really happening by giving numbers +1 and -1, would still result in a violation of Bell's inequalities, albeit not to the maximum extent of +4. The reason for this is that some of the participants would believe case C to be taking place, rather than suppose that one of the girls is mistaken.

*4.3. The violation of Bell's inequalities in language*

Before we start to analyze the violation of Bell's inequalities we wish to make the following remark. Suppose that we have the following product equalities:

E(*e*, *p*, *f*) = E(*e*, *p*) E(*p*, *f*)
E(*e*, *p*, *g*) = E(*e*, *p*) E(*p*, *g*)
E(*g*, *p*, *f*) = E(*g*, *p*) E(*p*, *f*)
E(*g*, *p*, *g*) = E(*g*, *p*) E(*p*, *g*)

In this case we have:

|E(*e*, *p*, *f*) - E(*e*, *p*, *g*)| + | E(*g*, *p*, *f*) + E(*g*, *p*, *g*)|

= | E(*e*, *p*) E(*p*, *f*) - E(*e*, *p*) E(*p*, *g*)| + | E(*g*, *p*) E(*p*, *f*) + E(*g*, *p*) E(*p*, *g*)|

= | E(*e*, *p*) (E(*p*, *f*) - E(*p*, *g*))| + | E(*g*, *p*) (E(*p*, *f*) + E(*p*, *g*))|

= | E(*e*, *p*) | | E(*p*, *f*) - E(*p*, *g*) | + | E(*g*, *p*) | | E(*p*, *f*) + E(*p*, *g*) |

The fact that E(*e*, *p*) = +1 or E(*e*, *p*) = -1 implies that | E(*e*, *p*) | = +1, and similarly | E(*g*, *p*) | = +1, hence:

= | E(*p*, *f*) - E(*p*, *g*) | + | E(*p*, *f*) + E(*p*, *g*) |

The fact that E(*p*, *f*) = +1 or E(*p*, *f*) = -1, and E(*p*, *g*) = + 1 or E(*p*, *g*) = -1, yields
| E(*p*, *f*) - E(*p*, *g*) | = 0 or | E(*p*, *f*) - E(*p*, *g*) | = +2, and also | E(*p*, *f*) + E(*p*, *g*) | = 0 or | E(*p*, *f*) + E(*p*, *g*) | = +2, and clearly if | E(*p*, *f*) - E(*p*, *g*) | = 0 then | E(*p*, *f*) + E(*p*, *g*) | = +2, while if | E(*p*, *f*) - E(*p*, *g*) | = +2 then | E(*p*, *f*) + E(*p*, *g*) | = 0, which proves that

| E(*p*, *f*) - E(*p*, *g*) | + | E(*p*, *f*) + E(*p*, *g*) | = +2

This proves that in this case Bell's inequalities are never violated. This means that any violation of Bell's inequalities requires a violation of at least one of the product equalities. In our example of the two pets eating their favourite food, we have

$E(e, p, f) \neq E(e, p) E(p, f)$

Indeed, $E(e, p) = +1$, because Roller is the one who is eating (because of what Amy has seen), and $E(p, f) = +1$, because it is Eukanuba that is being eaten (because of what Carol smells), but $E(e, p, f) = -1$, because the overall context makes it very probable that one of the girls is mistaken.

What does all this show?

(i) The origin of the violation is the fact that in the sentence *The pet eats the food* the concepts *pet* and *food* are entangled. More concretely, this means that when the concept *cat* collapses to *Felix*, the concept *food* collapses to *the food that Felix is eating*. We formulated the overall contexts in such a way that this collapse is well defined, so that if *pet* collapses to *Felix*, then *food* collapses to *Eukanuba*. In the overall contexts of the world at large, this entanglement will not be so neatly defined. Even so, the fact is that if *pet* collapses to a specific *pet*, the *food* collapses to a specific *food*, namely the *food* that this specific *pet eats*. This underlying mechanism of entanglement is the mechanism that carries the meaning of the sentence *The pet eats the food*, and it is this which makes this sentence different from *just the bag of words* {pet, eat, food}. That the tensor product can be used to describe this entanglement, exactly as it is done in quantum mechanics, is shown explicitly in (Aerts and Gabora in press/ab), where the full quantum mathematical description in Hilbert space is worked out for the sentence *The cat eats the food*.

(ii) The violation also requires aspects of language that are not purely logical. It does play a role that we have constructed a situation where the meaning of a particular combination of sentences allows concluding that one of the two girls must be mistaken. This would not be the case if we reduced the situation to a set of logical propositions. In this respect it is worth noting the following: Bell's inequalities are violated because in case $(e, p, f)$ it is plausible that one of the two girls is mistaken, and that it is not Roller who is eating cat food. Of course, also in cases $(e, p, g)$ and $(g, p, f)$ it is quite possible that one of the girls is mistaken. But language functions in such a way that this possibility will be much less taken into account, because there is no reason to believe that one of the girls is mistaken in these situations, whereas there is in situation $(e, p, f)$, which is contradictory, given the assumption that Roller never eats cat food.

(iii) We might be tempted to believe, taking into account our remark in (ii), that violations of Bell's inequalities in language are strictly linked to contradictory situations. This is not true, however. The contradiction per se is of no importance. What is important is that the effect of context $e$ on sentence $p$ (and more specifically on the concept *pet* as part of sentence $p$) depends on whether it is context $f$ that affects sentence $p$ (and more specifically the concept *food* as part of sentence $p$) or context $g$ that affects sentence $p$ (and more specifically the concept *food* as part of sentence $p$). If this is the case, we have

$E(e, p, f) \neq E(e, p) E(p, f)$

with a violation of Bell's inequalities being possible. Our example contains a contradiction (one of the two girls is mistaken) only for the sake of making a clear case without having to perform real experiments on subjects. Although it is obvious that the effect may well be less strong in other contexts, it will still be there, so that we can say that any such other contexts will still give rise to a violation of Bell's inequalities.

We had presented in earlier work a situation where Bell's inequalities are violated in cognition (Aerts, Aerts, Broekaert et al. 2000). However the situation considered there introduced conceptual as well as physical contexts, and hence cannot be reduced easily to a purely linguistic situation, as the one considered in the present article.

*4.4. The implications for Evolutionary Epistemology*

Section 3 outlined the presence of quantum structures in decision processes and concept structures. Section 4 presented an explicit violation of Bell's inequalities in language, demonstrating the presence of non-Kolmogorovian probability structure in language. Section 2 pointed out that the mathematical models of Markov chains, used to describe the neo-Darwinian mechanism of evolution through variation and selection, is built upon a Kolmogorovian probability model. Technically, this follows from the fact that Markov processes are formulated using a probability structure that is defined on a σ-algebra of events[13], and this σ-algebra of events makes the probability model Kolmogorovian. Neo-Darwinian evolution is modelled in this way, because this is the natural way to model a mechanism of variation and selection on a *set of actualised entities*, be they phenotypes or genes or other units of evolution. The fact that Bell's inequalities are violated means that the probability structure involved is non-Kolmogorovian, and hence that evolution cannot be modelled using Markov processes. In Aerts, Czachor, Gabora, et al. (2003), we designed a global mathematical evolution model that incorporates non-Kolmogorovian probabilities. We are currently elaborating this approach by focusing on non-Kolmogorovian models for a replicator dynamics. We are well aware of the highly abstract nature of our criticism of the neo-Darwinian evolution scheme as it stands now. In this sense, it is not obvious whether it can lead to new insights or solutions to some of the well-known problems regarding neo-Darwinian evolution. In view of this, we will outline what we expect might result once we have elaborated much more sophisticated models. Our experience with the effects of quantum models when used in the realm of the micro-world allows us to formulate these expectations in some detail. Of course, the correctness of this speculation depends on how complete the quantum structure of which we have detected aspects is present in the macro-world, more specifically in its psycho-cognitive layer.

A first aspect of *quantum change* that we should mention is the following. There are two types of change: the one is called *collapse*, defined as the change under the influence of a specific and nearby context. The other change is called *dynamical change*, which is defined as a change, not under the influence of a specific nearby context, but rather as the consequence of the global interaction with the rest of the universe. A second aspect to mention is that there are also two types of state with respect to a specific context of the entities involved in evolution. There is a type of state called an *eigenstate*, by which is meant a state where the entity is actualised with respect to this context; more specifically, it is the state the entity is in after it has undergone a collapse type of change under the influence of this context. The second type of state is called a *superposition state*. This is a state that is not actualised with respect to this context, hence it is a potentiality state (Aerts and D'Hooghe in press). If it is in such a superposition state, and under the influence of this context, the entity will collapse to one of the eigenstates with a certain probability.

Let us give an example of the type of situation that we think can be described using this formalism. Suppose we consider again the cat food Eukanuba of section 4.2. Of course, there are many other brands of cat food on the market, for instance Purina, Sophistacat, Bil-Jac, Iams, Whiskas and Friskies. All of these existing cat food brands are competing on the same North-American market. This is a good example of where in a rather obvious way the neo-Darwinian scheme can be applied. Indeed, some of the brands can be fitter than the others, and if the difference in fitness is too big, the non-fit ones will disappear from the market. All these seven brands are *actualized entities* with respect to the context, *i.e.* the North-American market. But what about the situation where a new company plans to market a new brand of cat food? Probably there is a particular time-frame in which the new brand exists only in the planning phase: the plans to market the brand are there, and become more and more concrete, but the brand is not yet an *actualized entity* in the market. In our quantum formalism, we will describe this new brand as being in a *superposition state* with respect to the context, which is the North-American market, while an actualized brand is in an *eigenstate* with respect to this context. Of course, in this superposition state (potentiality state), the brand will already be subject to evolution, as the plans evolve. But they evolve in the realms of the interacting minds of the designers, engineers, chemists and marketers that are in the planning of the new brand. The effect of the market on this superposition state is a collapse-type of change; indeed, slowly but definitely the plans for the new brand of cat food change towards an actualization of a new brand on the market. Following its actualization, it will then openly compete with the other brands, which have already been actualized.

---

[13] A σ-algebra is a Boolean algebra that is σ-complete, meaning that denumerable infinite conjunctions and disjunctions of events are allowed, while in a Boolean algebra only finite conjunctions and disjunctions are allowed.

      Now there are some interesting questions to ask. For example, is there already competition in the superposition phase of evolution? Certainly there is. Indeed, to see this clearly, we just have to consider a situation that is somewhat more complex. Let us suppose that there are two companies considering the introduction of a new brand of cat food. If they know of each other's intentions, there certainly will already be competition in the planning or superposition phase. And, of course, there will also be competition between the not yet actualized brands - the brands in the planning phase, hence in a superposition state - and the already actualized brands, in an eigenstate. It might well be that some of the actualized brands temporarily withdraw to the superposition state again, to allow for plans to be made for adjusting to the newcomers on the market. Our expectation is that all of this possible dynamics will be readily described by a quantum evolution scheme.

      The final remark we wish to make concerning EE is the most speculative one, but it is too interesting to be left out. We believe that, with regard to EE, things should be seen in exactly the opposite way. Biological evolution should not be taken as a metaphor for epistemological and conceptual change. It is epistemological and conceptual change that should be taken as a metaphor for biological evolution. We do believe that the change that we readily see happening in front of our eyes in the conceptual, cultural, psychological and social realities around us – let us call this EE-type of evolution – is the more general one, with biological evolution being the more specific one, in some sense a special case of the EE-type evolution. That is why we believe that there are quantum aspects to biological evolution too, although they are much less obvious than the ones we have identified in EE-type of evolution. Let us consider for a moment in this respect one of the biggest problems encountered in biological evolution: the gaps in the fossil records. Now if we were to suppose that biological evolution is quantum-like, we might argue that in this case only actualized life forms, hence life forms in collapsed states with respect to the context – the earth's environment –, give rise to fossils, and that a not yet detected undercurrent reality exists, where life forms in superposition states evolve and interact with each other, without ever giving rise to fossils. In this undercurrent of reality, life forms would thus exist in potentiality states, albeit potentiality states that could collapse to an actualized life form, but not necessarily so. These non actualized life-forms evolve under a similar type of evolution as the one that governs epistemological and conceptual entities, which means that they could

give rise to new non-actualized life forms, without leaving traces in the context which is the earth's environment. This would explain the gaps in the fossil records and the sudden appearance of new life forms as a natural phenomenon, the equivalent of the sudden appearance of a new brand of cat food on the North-American market, after the brand has evolved in the planning phase within the human minds that are engaged in its planning.

**5. Conclusion**
The neo-Darwinian mechanism of evolution formalized by means of Markov processes on the level of the phenotypes entails an underlying Kolmogorovian probability model. This Kolmogorovian structure of the probability model is not just due to an arbitrary choice for a Markov chain realization of the neo-Darwinian evolution process. The neo-Darwinian accent on *variation* and *selection* as basic mechanisms leads to this Markov formalization, because variation and selection are considered to function on a set of *actualized units of evolution*. We argue in this article that this neo-Darwinian mechanism falls short of describing evolution, because of its limitation to a Markovian process structure, whenever non-Kolmogorovian probability structures are present. We describe the different situations in which we have proved the presence of non-Kolmogorovian probability structure in former research work, more specifically in human decision processes, in how concepts and combinations of concepts are structured. In this article we show the presence of non-Kolmogorovian probability structure in language. Decision processes, concepts and their combinations, and language, are three situations that appertain to the field that is also considered by EE. This means that our finding shows that if EE is interpreted narrowly, and hence attempt to implement the neo-Darwinian mechanism to epistemology, it will stumble upon the problem of the presence of non-Kolmogorovian probability structures in some of the situations of the evolution that it attempts to model. Hence this article is an argumentation for the broad interpretation of EE. More speculatively, we have suggested that it would be preferable to regard epistemological and conceptual evolution (or rather its generalisation, which we have called EE-type of evolution – see section 4.4.) as the more general one, and biological evolution as a special case of it, in the same way as for example quantum mechanics is the more general type of mechanics, and classical mechanics is a special case of it. If this line of reasoning is followed with respect to biological evolution, so that also biological evolution would contain quantum aspects (as we believe it does, and we are working on a project to corroborate this), we can sketch a speculative scenario that would provide a natural explanation for the well-known problem in biological evolution of the gap in the fossil records and the sudden appearance of new life forms (section 4.4.).

**References**

Accardi, Luigi. 1984. "The probabilistic roots of the quantum mechanical paradoxes". In: Diner, Simon; Fargue, Daniel; Lochak, George et al. (eds), *The wave-particle dualism* 47-55. Dordrecht: Kluwer Academic.

Aerts, Diederik. 1982. "Example of a macroscopical situation that violates Bell inequalities". *Lettere al Nuovo Cimento* 34: 107-111.

Aerts, Diederik. 1986. "A possible explanation for the probabilities of quantum mechanics". *Journal of Mathematical Physics* 27: 202-210.

Aerts, Diederik. 1995. "Quantum structures: an attempt to explain their appearance in nature". *International Journal of Theoretical Physics* 34: 1165-1186.

Aerts, Diederik; and Aerts, Sven. 1994. "Applications of quantum statistics in psychological studies of decision processes". *Foundations of Science* 1: 85-97.

Aerts, Diederik; and Aerts, Sven. 1997. "Application of quantum statistics in psychological studies of decision processes". In: van Fraassen, Bas C. (ed.), *Topics in the foundation of statistics* 92-104. Dordrecht: Kluwer Academic.

Aerts, Diederik; Aerts, Sven; Broekaert, Jan et al. 2000. "The violation of Bell inequalities in the macroworld". *Foundations of Physics* 30: 1387-1414.

Aerts, Diederik; Apostel, Leo; De Moor, Bart et al. 1995. *Perspectives on the world, an interdisciplinary reflection.* Brussels: VUBPress.

Aerts, Diederik; and Czachor, Marek. 2004. "Quantum aspects of semantic analysis and symbolic artificial intelligence". *Journal of Physics A-Mathematical and General* 37: L123-L132.

Aerts, Diederik; Czachor, Marek; Gabora, Liane et al. 2003. "Quantum morphogenesis: a variation on Thom's catastrophe theory". *Physical Review E* 67: 051926-1-051926-13.

Aerts, Diederik; and D'Hooghe, Bart. in press. "Potentiality states: quantum versus classical emergence". *Foundations of Science.*

Aerts, Diederik; and Gabora, Liane. in press a. "A theory of concepts and their combinations I: The structure of the sets of contexts and properties". *Kybernetes*.

Aerts, Diederik; and Gabora, Liane. in press b. "A theory of concepts and their combinations II: A Hilbert space representation". *Kybernetes*.

Apostel, Leo. 1995. "The unfinished symphony: positions, agreements, disagreements and gaps". In: The worldviews group (ed.), *Perspectives on the world: an interdisciplinary reflection* 219-239. Brussel: VUBPress.

Aspect, Alain. 1999. "Bell's inequality test: more ideal than ever". *Nature* 398: 189-190.

Bell, John S. 1964. "On the Einstein-Podolski-Rosen paradox". *Physics* 1: 195-200.

Bell, John S. 1966. "On the problem of hidden variables in quantum mechanics". *Reviews of Modern Physics* 38: 447-452.

Blei, David M.; Ng, Andrew Y.; and Jordan, Michael I. 2003. "Latent Dirichlet allocation". *Journal of Machine Learning Research* 3: 993-1022.

Clauser, John F.; and Shimony, Abner. 1978. "Bell's theorem: experimental tests and implications". *Reports on Progress in Physics* 41:1881-1926.

Czachor, Marek. 1992. "On classical models of spin". *Foundations of Physics Letters* 3: 249-264.

Czachor, Marek; and Naudts, Jan. 2002. "Thermostatistics based on Kolmogorov-Nagumo averages: unifying framework for extensive and nonextensive generalizations". *Physics Letters A* 298: 369-374.

Deerwester, Scott; Dumais, Susan T.; Furnas, George W. et al. 1990. "Indexing by Latent Semantic Analysis". *Journal of the American Society of Information Science* 41: 391-407.

Doob, Joseph L. 1953. *Stochastic processes*. New York: John Wiley and Sons.

Einstein, Albert; Podolsky, Boris; and Rosen, Nathan. 1935. "Can quantum-mechanical description of physical reality be considered complete". *Physical Review* 47: 777-780.

Fodor, Jerry. 1994. "Concepts: a potboiler". *Cognition* 50: 95-113.

Foulis, David J.; and Randall, Charles H. 1972. "Operational statistics I". *Journal of Mathematical Physics* 13: 1667-1675.

Gabora, Liane; and Aerts, Diederik. 2002a. "Contextualizing concepts". *Proceedings of the 15th international FLAIRS conference, special track: Categorization and concept representation: models and implications* 148-152. Pensacola Florida, May 14-17, American Association for Artificial Intelligence.

Gabora, Liane; and Aerts, Diederik. 2002b. "Contextualizing concepts using a mathematical generalization of the quantum formalism". *Journal of Experimental and Theoretical Artificial Intelligence* 14: 327-358.

Griffiths, Thomas L.; and Steyvers, Mark. 2004. "Finding scientific topics". *Proceedings of the National Academy of Sciences* 101: 5228-523.

Gudder, Stanley P. 1988. *Quantum probability*. San Diego: Academic Press.

Gudder, Stanley P. 1968. "Hidden variables in quantum mechanics reconsidered". *Reviews of Modern Physics* 40: 229-231.

Hampton, James. 1997. "Conceptual combination". In: Lamberts, Koen; and Shanks, David (eds), *Knowledge, concepts, and categories* 133-159. Hove: Psychology Press.

Heit, Evan; and Barsalou, Lawrence. 1996. "The instantiation principle in natural language categories". *Memory* 4: 413-451.

Hofmann, Thomas. 1999. "Probabilistic latent semantic analysis". In: Laskey, Kathryn B.; and Prade, Henri (eds), *Proceedings of the fifteenth conference on uncertainty in artificial intelligence UAI'99* 289-296. San Francisco: Morgan Kaufmann.

Hoppensteadt, Frank C. 1982. *Mathematical methods of population biology*. Cambridge: University Press.

Jauch, Jozef M.; and Piron, Constantin. 1963. "Can hidden variables be excluded in quantum mechanics". *Helvetica Physica Acta* 38: 827-837.

Kamp, Hans; and Partee, Barbara. 1995. "Prototype theory and compositionality". *Cognition* 57: 129-191.

Kochen, Simon; and Specker, Ernst P. 1967. "The problem of hidden variables in quantum mechanics". *Journal of Mathematical Mechanics* 17: 59-87.

Kolmogorov, Andrey N. 1950. *Foundations of the theory of probability*. New York: Chelsea.

Komatsu, Lloyd K. 1992. "Recent views on conceptual structure". *Psychological Bulletin* 112: 500-526.

Landauer, Thomas K. 2002. "On the computational basis of learning and cognition: arguments from LSA". In: Ross, Brian H. (ed.), *Psychology of learning and motivation* 41: 43-84. New York: Academic Press.

Landauer, Thomas K.; and Dumais, Susan T. 1997. "A solution of Plato's problem: The Latent Semantic Analysis theory of the acquisition, induction, and representation of knowledge". *Psychological Review* 104: 211-240.

Landauer, Thomas K.; Foltz, Peter W.; and Laham, Darrell. 1998. "Introduction to Latent Semantic Analysis". *Discourse Processes* 25: 259-284.

Landauer, Thomas K.; Laham, Darrell; and Foltz, Peter W. 1998. "Learning human-like knowledge by Singular Value Decomposition: a progress report". *Advances in Neural Information Processing Systems* 10: 45-51.

Lund, Kevin; and Burgess, Curt. 1996. "Producing high-dimensional semantic spaces from lexical co-occurence". *Behavior Research Methods, Instruments and Computers* 28: 203-208.

Markov, Andrei A. 1906. "Rasprostranenie zakona bol'shih chisel na velichiny, zavisyaschie drug ot druga". *Izvestiya Fiziko-matematicheskogo obschestva pri Kazanskom universitete, 2-ya seriya* 15: 135-156.

Montemurro, Marcello A. 2001. "Beyond Zipf-Mandelbrot law in quantitative linguistics". *Physica A* 300: 563-578.

Neumann, John von. 1932. *Mathematische grundlagen der quanten-mechanik*. Berlin: Springer-Verlag.

Nosofsky, Robert. 1988. "Exemplar-based accounts of relations between classification, recognition, and typicality". *Journal of Experimental Psychology: Learning, Memory, and Cognition* 14: 700-708.

Nosofsky, Robert. 1992. "Exemplars, prototypes, and similarity rules". In: Healy, Alice F.; Kosslyn, Stephen M.; and Shiffrin, Richard M. (eds), *From learning theory to connectionist theory: essays in honor of William K. Estes* 149-168. Hillsdale NJ: Lawrence Erlbaum.

Osherson, Daniel N.; and Smith, Edward E. 1981. "On the adequacy of prototype theory as a theory of concepts". *Cognition* 9: 35-58.

Osherson, Daniel N.; and Smith, Edward E. 1982. "Gradedness and conceptual combination", *Cognition* 9: 35-58.

Pitowsky, Itamar. 1989. *Quantum probability - quantum logic*, Lecture Notes in Physics 321. Berlin: Springer-Verlag.

Plate, Tony A. 2003. *Holographic reduced representation: distributed representation for cognitive structures*. Stanford: CSLI Publications.

Rips, Lance J. 1995. "The current status of research on concept combination". *Mind and Language* 10: 72-104.

Rosch, Eleanor. 1975. "Cognitive representations of semantic categories". *Journal of Experimental Psychology: General* 104: 192-232.

Rosch, Eleanor; and Mervis, Carolyn B. 1975. "Family resemblances: studies in the internal structure of categories". *Cognitive Psychology* 7: 573-605.

Rosch, Eleanor. 1978. "Principles of categorization". In: Rosch, Eleanor; and Lloyd, Barbara B. (eds), *Cognition and categorization* 27-48. Hillsdale, NJ: Lawrence Erlbaum.

Rosch, Eleanor. 1983. "Prototype classification and logical classification: the two systems". In: Scholnick, Ellen K. (ed.), *New trends in conceptual representation: challenges to Piaget's theory?* 73-86. Hillsdale, NJ: Lawrence Erlbaum.

Smith, Edward E.; and Medin, Douglas L. 1981. *Categories and concepts*. Cambridge, MA: Harvard University Press.

Smolensky, Paul. 1990. "Tensor product variable binding and the representation of symbolic structures in connectionist systems", *Artificial Intelligence* 46: 159-216.

Widdows, Dominic; and Peters, Stanley. 2003. "Word vectors and quantum logic: experiments with negation and disjunction". In: *Proceedings of the eighth mathematics of language conference* 141-154. Bloomington, Indiana.

Zadeh, Lofti. 1982. "A note on prototype theory and fuzzy sets". *Cognition* 12: 291-297.